\documentclass{article}

\usepackage{graphics}
\usepackage{graphicx}
\usepackage{amsmath}
\usepackage{amssymb}
\usepackage{subfigure}
\begin{document}

\title{Characterizing neuromorphologic alterations with additive shape
functionals}

\author{M. S. Barbosa \and L. da F. Costa \and
  E. S. Bernardes \and G. Ramakers \and J. van Pelt}


\maketitle

\begin{abstract}
  The complexity of a neuronal cell shape is known to be related to its
  function. Specifically, among other indicators, a decreased complexity in
  the dendritic trees of cortical pyramidal neurons has been associated with
  mental retardation. In this paper we develop a procedure to address the
  characterization of morphological changes induced in cultured neurons by
  over-expressing a gene involved in mental retardation.  Measures associated
  with the multiscale connectivity, an additive image functional, are found to
  give a reasonable separation criterion between two categories of cells. One
  category consists of a control group and two transfected groups of neurons,
  and the other, a class of cat ganglionary cells. The reported framework also
  identified a trend towards lower complexity in one of the transfected
  groups. Such results establish the suggested measures as an effective
  descriptors of cell shape.
\end{abstract}

\section{Introduction}
\label{intro}

Classical investigations in neuroscience have added a large amount of
descriptive data to the knowledge of the mammalian neural system and
ultimately of the functioning of the brain. In recent years the use of
computer and algorithmical methods has been incorporated into the biological
sciences with the aim of achieving quantitative analysis and modelling. Being
no exception to this important tendency, mathematic-computational
investigations in neuroscience are still in an incipient stage, with many
challenging questions and potential for important results. While the effort to
model quantitatively the emergent features of a neural system or element
(e.g. a single cell) involves its own technical difficulties, neuroinformatics
has already led to important results regarding the interplay between the
morphology of dendrites and their firing patterns~\cite{Ooyen:2002}. One
difficult aspect implied by the use of computers in neuroanatomy is the
relatively limited availability of accurate spatial data of real dendritic
arborizations for diverse morphological classes. The alternative approach of
generating morphologically realistic neurons has been successfully carried out
by using a complementary approach based either on recursive~\cite{Ascoli:1999}
or topological~\cite{Pelt:1999,Costa:1999} algorithmic implementation.

The anatomic details of the neuronal cells have been shown to represent a
relevant feature, for phylogenic and ontogenic studies, in the diagnosis of
diseases and in investigating the interplay between form and function
\cite{LU010,LU020,Costa:1999}. With the availability on the Internet of
combined databases of virtual and real (histochemically marked) neurons, the
analysis or morphological characterization of newly acquired neuronal
information can now be the subject of systematic processing by computer
methods, provided suitable shape descriptors are identified and applied.

There is a current interest in many branches of Physical and, more recently,
Biological sciences to study in a quantitative way the geometrical outline of
structures, both static and in development, and from that knowledge to probe
further into the functioning and justification of those forms. Examples of
results of such a quest in biology ranges from the elucidation of the
Fibonacci spirals emerging in
sunflowers~\cite{Douady&Coulder:1992,Douady&Coulder:1993} to the fractal
characterization of the nature of many botanical plants~\cite{Lind:1990}.  For
over a century the search aimed at establishing a relationship between the
form and function of neuronal cells has challenged and stimulated researchers,
starting from the pioneering efforts of Ram{\'o}n y
Cajal~\cite{Cayal:1894}. While many diseases have been diagnosed based on the
visual perception of morphological alterations in general tissue morphology,
automated approaches remain promising subjects of research, with many
potentially important applications.

Although it is now generally accepted that dendritic morphology plays a
crucial role in the functioning of the neural cell and ultimately in the
behaviour of the neural
system~\cite{Ooyen:2002,Washington:2000,Uylings:2002,Devaud:2000,Brain:2003},
there is no deterministic way to choose the best quantitative descriptor of
the geometry or topology of single neurons.  In order to be particularly
useful, such descriptors should correlate in some way with the respective
performed functions and behaviour, allowing abnormalities and specific
divergences from a healthy state or development to be clearly identified. An
additional benefit of such an investigation of the physical basis of neuronal
shape and growth is the possibility for investigating, through modelling, the
relationship between neuronal shape and function.

The study of biological forms requires the selection of shape
descriptors or shape functionals that fulfil a set of
requests. First, measures are expected to be objective and fast (in
order to treat a representative number of cases), while retaining
comprehensiveness and being potentially discriminative.  In addition,
such measures should lead to meaningful biological interpretations. An
important theoretical issue is the relative degeneracy of a set of
measures, a concept that reflects the fact that many different forms
or shapes produces the same or approximated measures. It has been
experimentally shown~\cite{LU010,LU021,LU099} that multiscale analysis
tends to augment the resolution power of geometric descriptors and,
despite the increased computational demands, this procedure provides
useful additional information and characterization of the considered
objects.  The neuronal cells have, in general, a complex spatial
tree-like structure with many dendritic bifurcations leading to
efficient spacial coverage, amplifying the influence area and
optimally connecting the neuron with a neighbouring cell's dendrites or
axons. This structure has been shown to exhibit different levels of
partial fractality at different spatial
scales~\cite{MOR89,Coelho:1996,LU020}.

A recently reported~\cite{Raedt01,Raedt02} procedure for calculating
additive shape functionals, known as Minkowski functionals gathered in
a framework called Integral-Geometry Morphological Image Analysis,
MIA, has been successfully applied to many areas of research including
Statistical Physics, Cosmology and Material
Sciences~\cite{Santalo,Stoyan,Raedt01,Raedt02,Mecke}. A first attempt
to bring those morphological concepts into biological sciences,
targeting the efficient discrimination of two main types of the
domestic cat's ganglionary neuronal cells, has been reported
recently~\cite{Barbosa:2002}. The use of additive functionals is
implemented in a pixelwise approach leading to the possibility of
straightforwardly obtaining the above mentioned multiscale fractal
dimension for complementary shape description. Every additive
continuous and motion invariant functionals in the 2D Euclidean plane
can be expressed as a combination of three Minkowski functionals which
are proportional to a known geometric quantity, namely the metric area
and perimeter and the topological Euler connectivity number. This
completeness extends to higher dimensions and the geometrical
functionals multiplicity is always one plus the dimension of the
lattice generalized voxels (e.g 3 in 2D plane, 4 in 3D space, etc.).
While the theory of Integral geometry provides a sophisticated set of
results and formulae, the practical implementation of the procedures
in Image analysis is relatively simple and
efficient~\cite{Raedt01,Raedt02}. The basic ideas of this methodology
is described in section~\ref{AF} and a simple example is worked out to
outline the procedure.

In this paper we investigate the importance of the spatial
distribution of branching points in neuronal arborizations to provide
a discriminative and informative characterization of the neuronal
morphology, with special attention given to additive shape
functionals. This novel procedure involves mapping the neuronal image
onto a set of points representing its bifurcation pattern, which are
subsequently dilated in order to produce a multiscale
representation~\cite{TAD01}, as the Minkowski functionals of such a
set of points is recorded. The obtained results substantiate the
potential of the above procedure regarding its application to a
database of 3 categories of rat neuronal cells and one class of cat
ganglion cells. The results are in accordance with the biological
importance of the spatial distribution of
branches~\cite{MOR89,SHO53,KRAPP98,Ooyen:2002,Uylings:2002,Washington:2000}
and provide a discriminative measure for the morphological
characterization of the neurons, indicating subtle morphological
differences as a consequence of the considered gene transfectation in
the rat categories.  This suggests that those treatments may not
significantly alter the topological aspect of the branching cell form
as far as the considered measurements are concerned.

\section{Additive Functionals}\label{AF} 

Integral geometry algorithms have been successfully used to
characterize morphologically complex patterns where the precise
process of formation is not completely known and is a subject of
modelling, see for example~\cite{Mecke:1996}. The central procedure is
the calculation of intrinsic volumes or Minkowski functionals (or yet
{\it querrmassintegrals}), which are a generalization of the usual
determination of volume. They can be defined as related to both
differential and integral geometry setups.  

\begin{figure}[htb]
\begin{centering} \includegraphics[scale=.25,angle=-90]{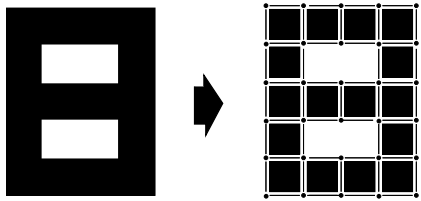}
\caption{The disjoint decomposition of a pattern $K$ (left) into a sum
  $\mathcal{P}$ (right) of disjoint interior elements. \label{fig:1}}
\end{centering}
\end{figure}

For the sake of a better explanation, we start with a description of
the practical aspect of the adopted procedure. The Minkowski
functionals of a body $K$ in the plane are proportional to its area
$A(K)$, perimeter $U(K)$ and the connectivity or Euler number
$\chi(K)$. The usual definition of the connectivity from algebraic
topology in two dimensions is the difference between the number of
connected $n_c $ components and the number of holes $n_h$,

\begin{equation} 
\chi(K)=n_c-n_h.  
\end{equation} 

So, if $K$ happens to be an image of the number 8 as in
Figure~\ref{fig:1}, its connectivity number would be $\chi(K
)=1-2=-1$. One instance where these functionals appear naturally is
while attempting to describe the change in area as the domain $K$, now
assumed to be convex, undergoes a dilation through a parallel set
process using a ball $B_r$ of radius r

\begin{equation}
\label{eq:dila}
A(K \oplus B_r)=A(K)+U(K)r+\pi r^2. 
\end{equation}   

Generalizing to higher dimensions, the change in hyper volume is given
by the Steiner formula~\cite{Santalo}

\begin{equation}
\label{eq_dilageral}
v^d(K\oplus B_r)=\sum_{\nu=0}^d \binom{d}{\nu}
W_{\nu}^{(d)}(K)r^{\nu}, 
\end{equation} 

where the coefficients $W_{\nu}^{(d)}$ are referred to as Minkowski
functionals. For instance in the plane ($d=2$),

\begin{equation}
\label{eq:coeffs}
  W_0^{(2)}(K)=A(K), W_1^{(2)}(K)=\frac{U(K)}{2}, W_2^{(2)}(K)=\chi(K)\pi.
\end{equation}

Despite the wealth of results and continuum formulae for obtaining
these functionals, it is useful to recur to the discrete nature of the
binary images we wish to analyse by looking at the distribution of
pixels in the planar lattice. On that level, by exploring the
additivity of the Minkowski functionals, their determination reduces to
counting the multiplicity of basic building blocks that compose the
object in a disjointed fashion. Figure \ref{fig:1} illustrates this
process. A pixel can be decomposed as a disjointed set of 4 vertices, 4
edges and one square. The same process can be applied to any object in
a lattice. The fundamental information needed here is a relationship
for the functionals of an open interior of a $n$-dimensional body $K$
which is embedded into a $d$-dimensional space

\begin{equation}\label{eq_open} 
W_{\nu}^{(d)}(\breve{K})=(-1)^{d+n+\nu}
W_{\nu}^{(d)}(K), \nu=0,\ldots, d.
\end{equation} 

As there is no overlap between these building blocks and by the
property of additivity of these functionals, for a body $P$ composed
of disjointed convex interior pieces $\breve{N}_m$, we may write

\begin{equation}\label{eq_whole} 
W_{\nu}^{(d)}(\mathcal{P})=\sum_{m}
W_{\nu}^{(d)}(\breve{N}_m)n_m(\mathcal{P}), \; \nu=0,\ldots,d.
\end{equation} 

Where $n_m(\mathcal{P})$ stands for the number of building elements of
each type $m$ occurring in the pattern $\mathcal{P}$. For a
two-dimensional space, which is our interest regarding the considered
neuronal images, we display in Table~\ref{tb_buildblocks} the value of
Minkowski functionals for the building elements in a square lattice of
pixels and their direct relation to familiar geometric quantities on
the plane. Using the information (with $a=1$) presented in
Table~\ref{tb_buildblocks} and equation \eqref{eq_whole} we have

\begin{equation}\label{eq_func} \quad
A(\mathcal{P})=n_2,\;U(\mathcal{P})=-4n_2+2n_1,\;\chi(\mathcal{P})=n_2-n_1+n_0.
\end{equation} 

Going back to Figure~\ref{fig:1}, we find, for this specific example,
as $n_2=16$, $n_1=47$ and $n_0=30$, that $A=16$, $U=30$ and $X=-1$.
So the procedure to calculate Minkowski functionals of a pattern $K$
has been reduced to the proper counting of the number of elementary
bodies of each type that compose a pixel (squares, edges and vertices)
involved in the make up of $\mathcal{P}$.  

\begin{table}[htb]
\caption{Minkowski functionals of open bodies $\breve{N}_m$ which
compose a
  pixel $K$: $\breve{P}$ ({\it vertex } ), $\breve{L}$ ({\it open edge}) and $\breve{Q}$ ({\it open square}). \label{tb_buildblocks}}
\begin{centering}
\begin{tabular}{l|l|l|l|l}\hline\hline
$m$ & $\breve{N}_m$  & $W_0^{(2)}=A(\breve{N}_m)$ & $W_1^{(2)}=\frac{1}{2}
U(\breve{N}_m)$ & $W_2^{(2)}=\pi\chi(\breve{N}_m)$ \\ \hline 
0 &  $\breve{P}$   & 0           & 0           & $\pi$    \\ \hline
1 &  $\breve{L}$   & 0           & $a$         & $-\pi$   \\ \hline
2 &  $\breve{Q}$   & $a^2$       & $-2a$       & $\pi$    \\ \hline\hline
\end{tabular}
\end{centering}
\end{table}

\section{Branching point patterns} The existence of profuse branching
structures in natural shapes provides evidence of the effectiveness of
such shapes in providing an interface with its environment as well as for
enabling connections. As pointed out in~\cite{Uylings:2002} many
conditions may interfere with the morphology of the neuronal tree
structure, for instance learning, 'enriched' environment, hormonal
fluctuations and levels of bioelectric activity.

The relationship between dendritic morphology and cell functioning,
alone or connected in networks, has drawn the attention of scientists
leading to a search for a set of measures that would as completely as
possible describe the neuronal shape~\cite{Hillman:1979}. For example,
the dependence of the dendritic diameter on the distance from soma,
the relationship between the dendritic diameter before a bifurcation
point and the diameter of the two daughters stemming from this point,
or the ratio of diameters of daughter dendrites. These are standard
examples of a global feature (the first) and two local properties (the
following two) one can specify/calculate in describing/analysing a
neuronal shape, real or virtual, see~\cite{Ascoli:1999}. There is a
vast recent literature describing shape analysis in general and
applied specifically to neuronal shape, see for
example~\cite{Uylings:2002,Kossel,KRAPP98} for a general dendritic
description, and~\cite{Coelho:1996} for multiscale fractality,
\cite{LU021} for wavelets and ~\cite{LU010} for bending energy applied
to neuromorphometry.

Since the pioneering work of Sholl~\cite{SHO53}, some authors have
mentioned specific situations where the well-known Sholl analysis
could lead to degenerate results, assigning to visually different
neurons the same (in statistical terms) descriptor values,
see~\cite{Kossel} for example. The method makes use of a reference
point at the soma and draws concentric circles through the dendritic
field, while counting the number of intersections within each
circle. While this process is easy to implement manually, which may
account for its popularity, it clearly fails for asymmetric
forms. While alternative approaches~\cite{Uylings:1989b} report
success in improving this setup, we take a different route, abandoning
the soma as a reference point.

We start our analysis by taking all bifurcation points and
constructing another image keeping the metric relation between these
points. Although there are automatic ways of extracting salient
points~\cite{LU078,LU009,costa:1997}, we implemented this task in a
semi-automated way, which is followed by a sequence of exact dilations
(permitted parallel set dilation~\cite{TAD01}) of the selected
points. Although primarily interested in the connectivity, all
Minkowski functionals are calculated by a routine that counts
efficiently the multiplicity of the building elements for each neuron
image at every radius of dilation as described in section~\ref{AF},
equation~\ref{eq_func}. The idea is to use the connectivity of the set
of points to capture, at each scale, the spatial relationship of the
neuron branching structure. As this measure is invariant to rotation
and translation, as well as to scaling, eventual variations of size
due to the relative stage of development can be disregarded.

We investigate the morphology of two different categories of cells. One
involves cat ganglion alpha cells, which constitutes a class by itself
but rather diversified in form,
see~\cite{Coelho:2002,Barbosa:2002}. The other includes foetal rat
cerebral cortex neurons cultured in a 2D tissue
system~\cite{Ramakers:1998}. This category is subdivided into three
differently treated classes consisting of a control group, a group of
neurons transfected with a gene OPHN1 which encodes for oligo-phrenin,
and a positive control group transfected with p190 RhoGAP.
            
\begin{figure*}[p] 
\begin{centering}
\begin{tabular}{lll}
\subfigure[OPHN1]{\includegraphics[scale=.3,angle=  0]{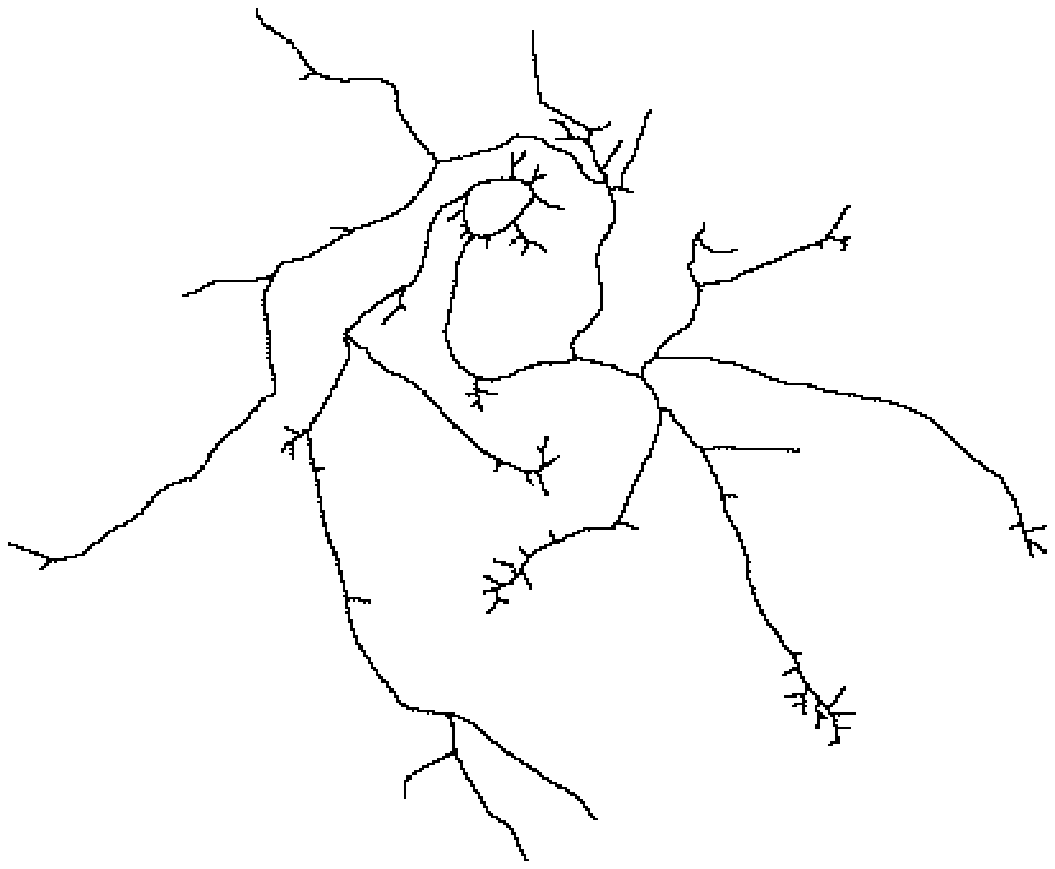}} &
\subfigure[]{\includegraphics[scale=.3,angle=0]{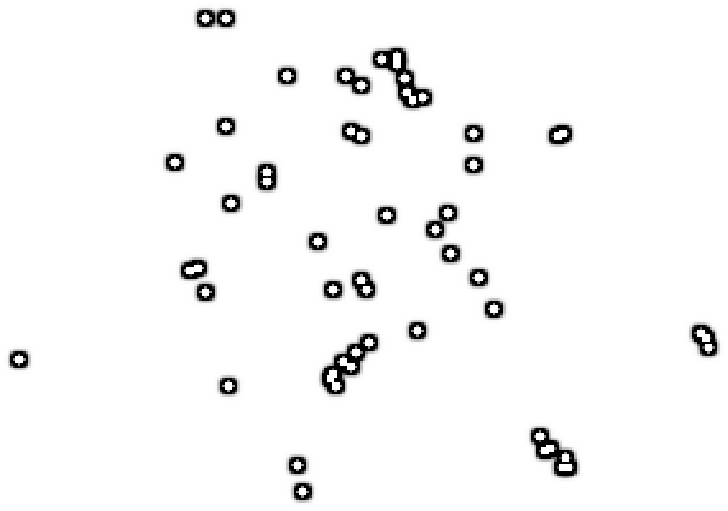}}&
\subfigure[]{\includegraphics[scale=.3,angle=0]{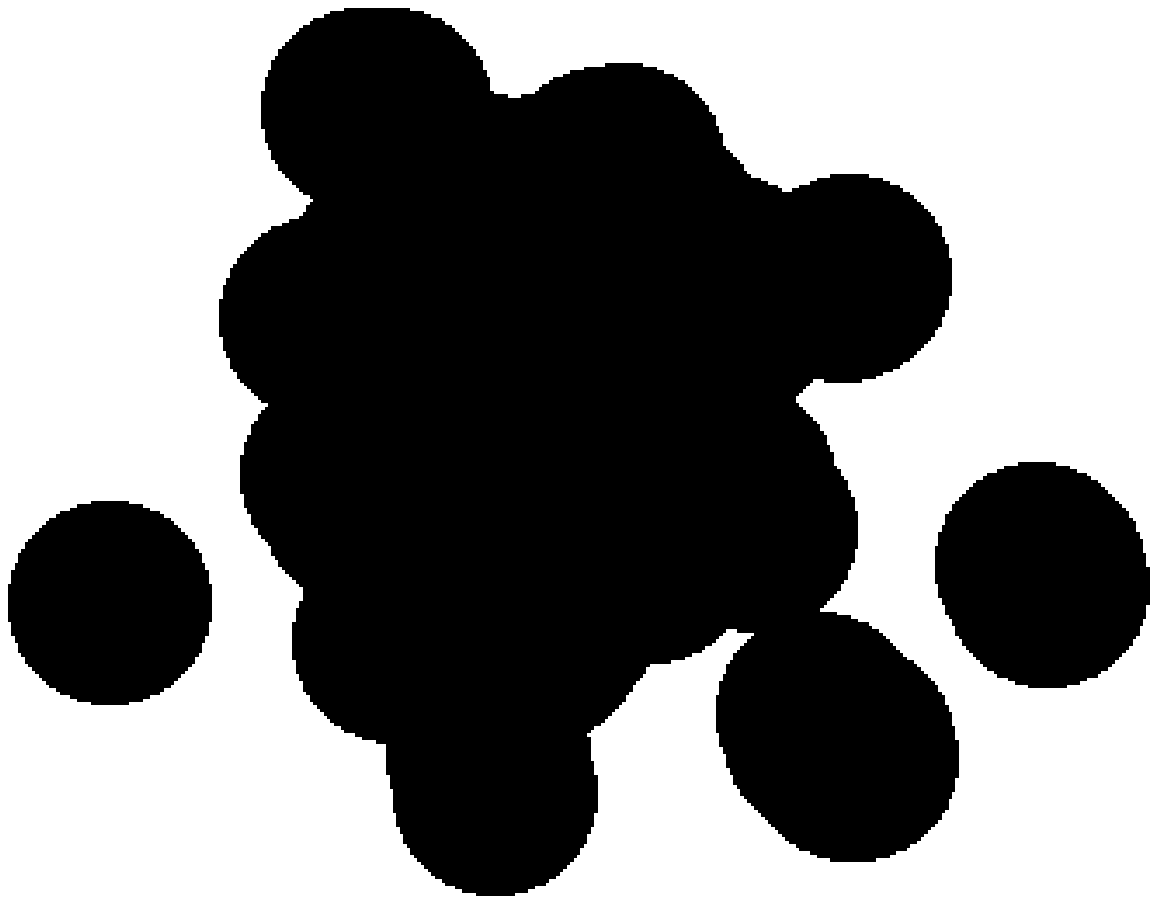}}\\
\subfigure[RhoGAP]{\includegraphics[scale=.25,angle=  0]{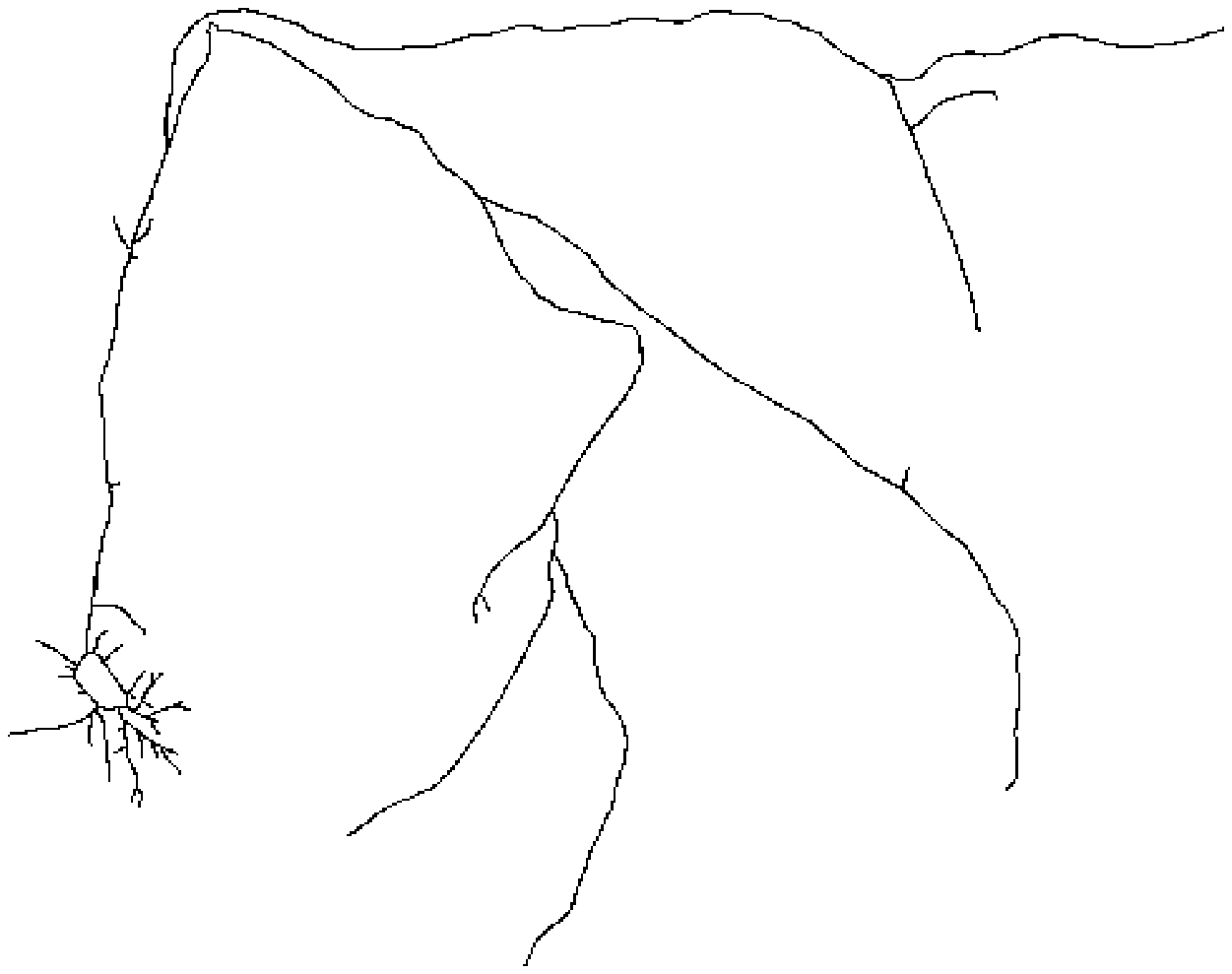}}&
\subfigure[]{\includegraphics[scale=.25,angle=0]{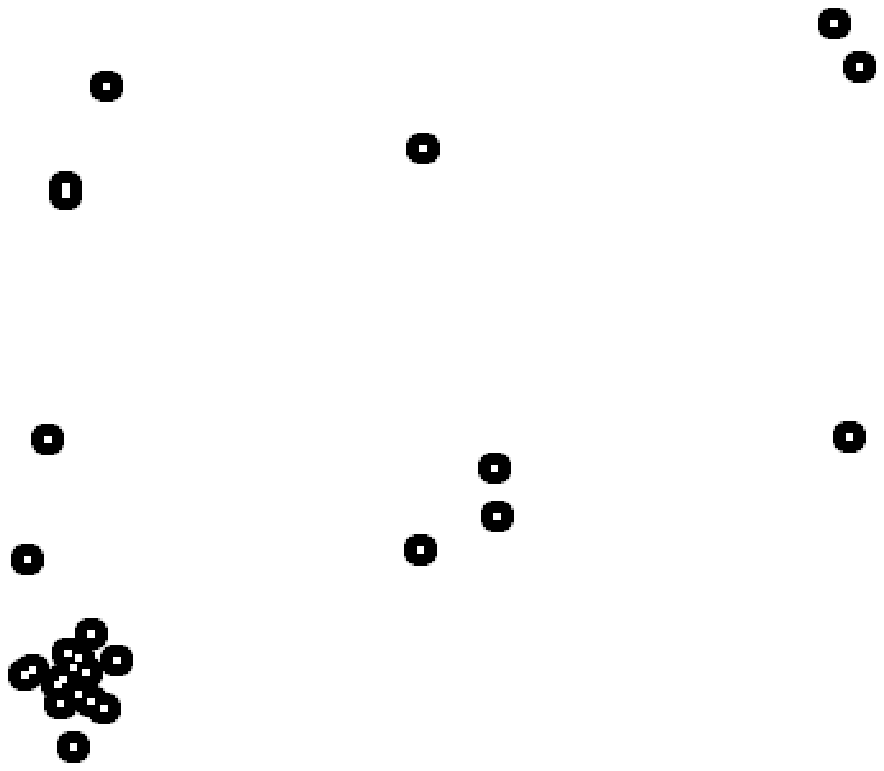}}&
\subfigure[]{\includegraphics[scale=.25,angle=0]{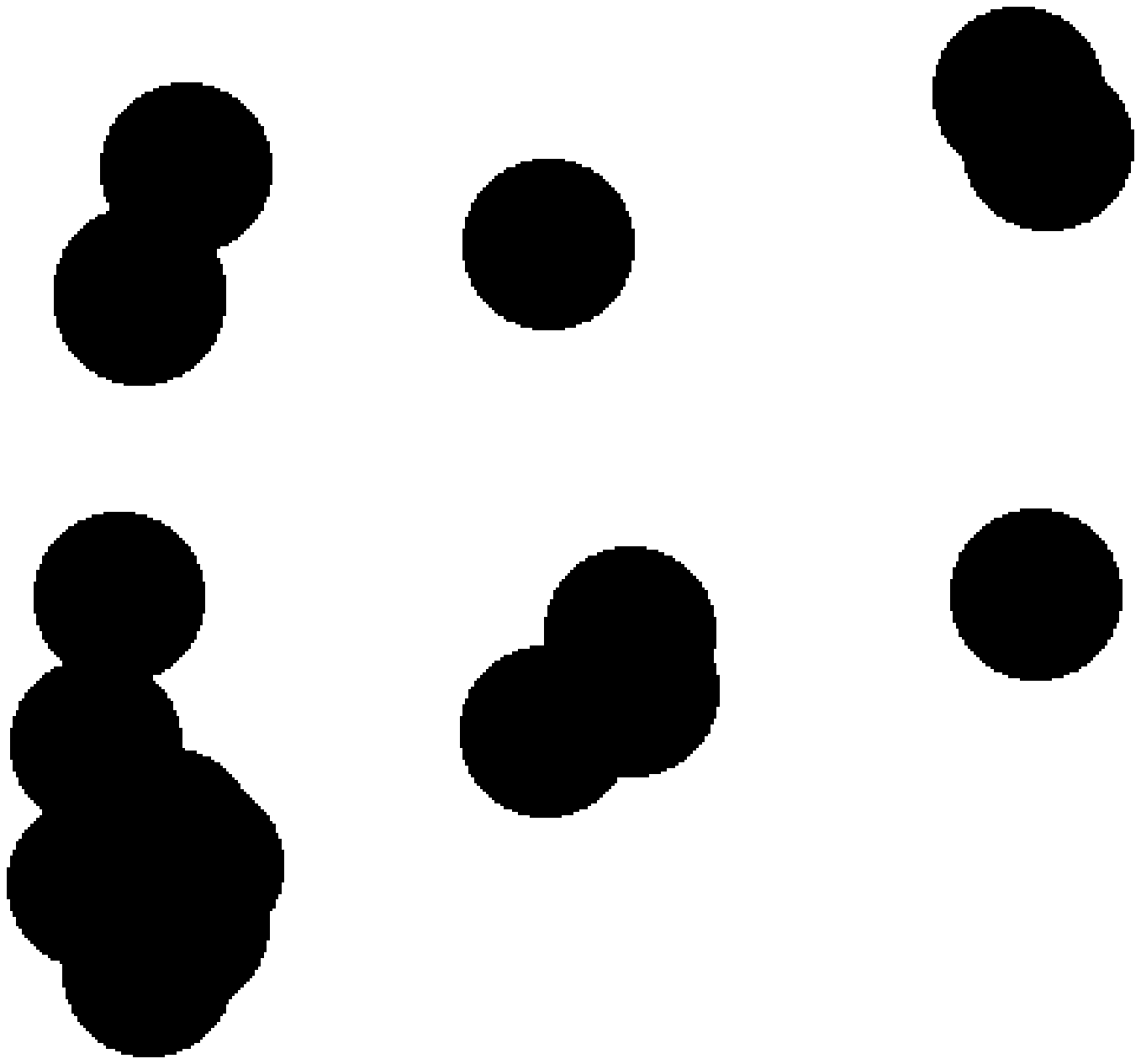}}\\
\multicolumn{3}{c}{
\subfigure[]{\includegraphics[scale=1.6,angle= -90]{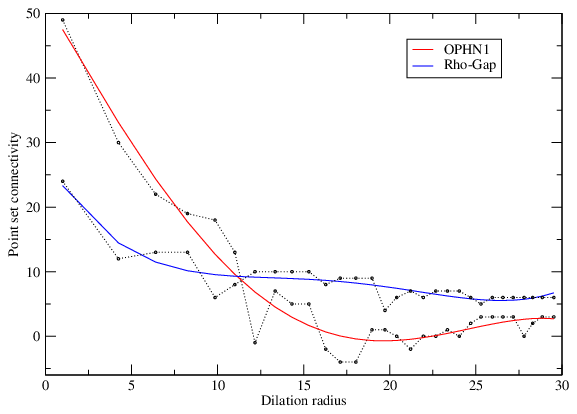}}}
\end{tabular}
\caption{Isolating morphological information relating to the neuronal branching
  structure while preserving the metric relationship among the branching
  points. Each bifurcating point undergoes an exact dilation while its
  connectivity is recorded as function of the scale parameter. 
  Prototypical transfected rat neuronal cell, a) and d). Extracted pattern of
  bifurcation points, b) and e). The end of the bifurcation point parallel set
  dilation procedure, c) and f). The resulting discriminating functional for
  both neuron types, g).\label{fig:2}}
\end{centering}
\end{figure*}

Figure~\ref{fig:2} shows examples of two treated rat neuronal cells,
illustrating the branching selection process adopted in this
work. Note that the neuron images have been rescaled in this picture,
but the final dilation radius in all processed samples are the same
and equal to 30 pixels.

\begin{figure*}[p]
\begin{centering}
\includegraphics[scale=0.45,angle=-90]{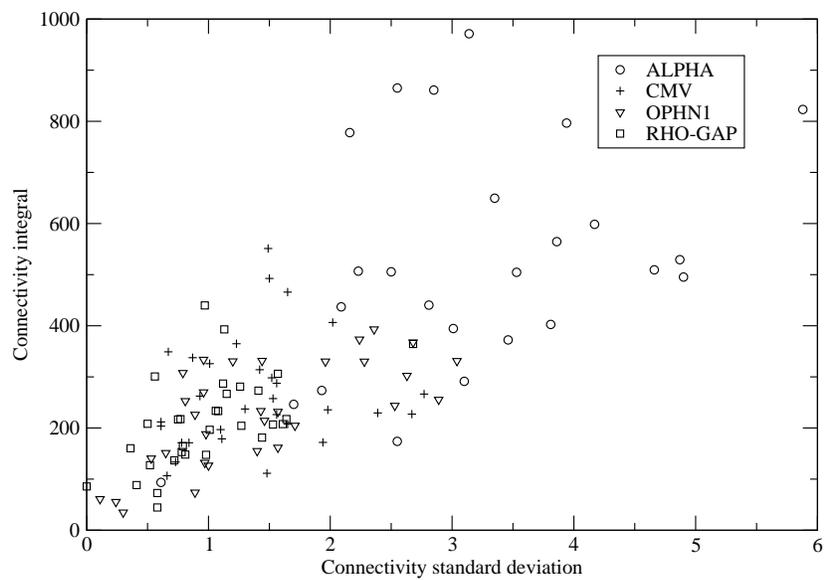}%
\caption{A discriminative feature space. Most of the Rho-Gap transfected
neurons went to the lower left corner of the plot, while the Alpha cat cells
are spread by themselves towards higher values. There is a fuzzy display of both
control and OPNH-1 samples. \label{fig:3}}
\end{centering}
\end{figure*} 

\begin{figure*}[p]
\begin{centering}
\includegraphics[scale=0.6,angle=0]{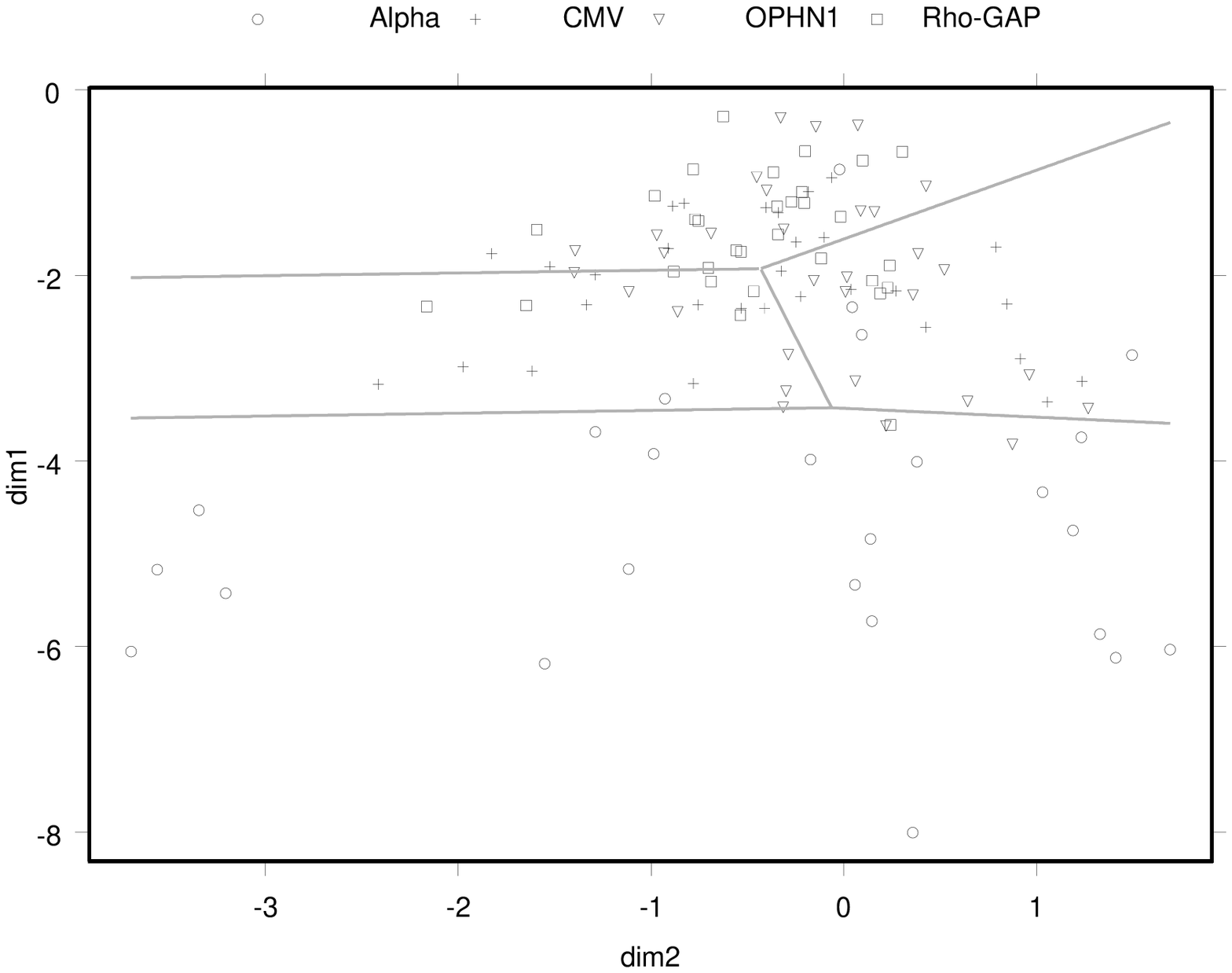}%
\caption{The discriminative power of the above feature space in depth. A
  discriminant analysis reveals a reasonable separation among ALPHA and Rho-GAP
  groups and is not conclusive for the others.\label{fig:4}}
\end{centering}
\end{figure*} 

It is important to comment on the complementary nature of the
framework proposed in this article with respect to more traditional
approaches reported in the literature.  When compared to the previous
Sholl methodologies, the measurements in our approach adopt
multiscale concepts so as to retain as much as possible the metric
and global structure, while avoiding the somewhat arbitrary use of the
soma as a general reference.  In order to express in a more objective
and effective way the morphological features of special interest,
namely the overall complexity of the neuronal pattern, our procedure
was designed from the start to concentrate on the bifurcation points,
which are used as primary data.  No attempt is made here to achieve
completeness in the shape description, or to use the considered
measures as features to be incorporated in an algorithm to grow
artificial neurons.  However, the reported measurements can be useful
for statistical validation of such artificial data, in the sense that
both the simulated and original neuronal structures should lead to the
same probability densities.

\section{Results} 
For each cell we calculate multiscale curves of connectivity as shown in
Figure~\ref{fig:2}. As discriminating measurements we take the area under the
interpolating curve (its integral) and the standard deviation of the
difference between that curve and the original data points. The former can
provide information about the overall structure, while the latter
characterized the finer details. Having performed similar analysis for area
and perimeter functionals, as well as for the multiscale fractality as derived
from the area data, we decided to focus on the multiscale connectivity of the
branching points set as a descriptor for its relationship to the spatial
distribution of branching points and its above average discriminating power.

Our primary concern here is to investigate the effectiveness of the
procedure adopted to quantify the subtle morphological aspects that
may characterize different cell types while also expressing a
biologically relevant attribution of the cell form, namely the
distribution of bifurcation points. In this regard, our results, shown
by Figure~\ref{fig:3}, led to a well-defined separation of the alpha
cells in a feature space defined by the integral of the multi scale
connectivity and its standard deviation.

Although the three differently treated rat cells are not clearly
distinguishable, one group of them, namely the one containing neurons
which show over expression of RHO-GAP following transfectation, clearly
tends to populate the lower left corner of the feature space. Yet,
while high dispersion is shown by the rats sample cells, including
the control group which is as much diversified in form as the treated
groups, the statistical tendency towards complexity agrees with a
previous analysis which focused on the multiscale fractal dimension of
this group of cells\cite{LU101}.

Figure~\ref{fig:4} shows a scatter plot defined by two canonical
discriminant functions (their canonical coefficients are displayed in
Table~\ref{tab:CC}) for the above selected shape descriptors. The
separability in groups is optimized (see Table~\ref{tab:MD} for a
quantitative index, the Mahalanobis distance) in such an analysis,
leading to a somewhat improved visibility of the above mentioned
distinction of the Rho-GAP over expressing set of cells from the
others. Nonetheless the other differently treated group mixes up
strongly with the control group, as can be verified from
Table~\ref{tab:Plug} which presents the plug-in (also called
confusion matrix) classification results. As can be seen from this
statistical analysis the alpha cat cell is not trivially
distinguishable from the others, as shown by the overlaps in both the
feature and canonical discriminant scatter plots. While the
effectiveness of use of Minkowski functionals (considering the whole
cell) has been previously reported~\cite{Barbosa:2002}, the framework
developed here is meant to emphasize the distribution of the
bifurcation points, in a way similar to that in Sholl's analysis, on
the cell morphology. 

\begin{table}
\caption{Canonical Coefficients}
\begin{tabular}{l|l|l}\hline\hline
   &        dim1       & dim2\\ \hline
Connectivity Integral &-0.893114489  &1.08447785\\ \hline
Connectivity Standard Deviation  &-0.003344825 &-0.00731277\\ \hline\hline
\end{tabular}
\label{tab:CC}
\end{table}

\begin{table}
\caption{Mahalanobis Distance}
\begin{centering}
\begin{tabular}{l|l|l|l|l}\hline\hline
        &   Alpha &     CMV   & OPHN1  &Rho-GAP\\ \hline
  Alpha &0.00 &6.08 &6.53 &8.97\\ \hline
    CMV &         &0.00 &0.10 &0.27\\ \hline
  OPHN1 &         &         &0.00 &0.31\\ \hline
Rho-GAP &         &         &         &0.00\\ \hline\hline
\end{tabular}
\label{tab:MD}
\end{centering}
\end{table}

\begin{table*}
\caption{Plug-in classification table}
\begin{centering}
\begin{tabular}{l|l|l|l|l|l|l}\hline\hline
        &Alpha &CMV &OPHN1 &Rho-GAP &    Error &Posterior.Error\\ \hline
  Alpha &   20 &  1 &    3 &      1 &0.20 &      0.25\\ \hline
    CMV &    0 &  9 &   10 &     11 &0.70 &      0.73\\ \hline
  OPHN1 &    2 &  6 &   10 &     13 &0.67 &      0.65\\ \hline
Rho-GAP &    1 &  6 &    5 &     19 &0.38 &      0.41\\ \hline\hline
\end{tabular}
\label{tab:Plug}
\end{centering}
\end{table*}

\section{Conclusions} 
This work describes how promising results have been obtained regarding a novel
procedure and measurements for classifying different types of neurons and to
reveal morphologically relevant attributes of different classes.  Although the
subtle morphological variations induced by the treatment of our samples are
masked by the statistical distributions and show only a small tendency towards
a reduction in complexity, the procedure shows clearly its discriminating
power when applied to a well characterized class of alpha cat ganglionary
cells. As the considered measurements were designed to express a particular
biological aspect of the neuronal geometry, namely the spatial relationship
among the branching points, we would expect that this specific trait is not
significantly affected by the gene transfectation process.  The procedure
reported in this paper can be generalised to process 3D Neurons, real or
virtual, and can be implemented straightforwardly. By extending the present
analysis to 3D we suspect that subtle information relating to neuron
complexity could be significantly improved.

\section{Acknowledgements}

The authors thank Regina C{\'e}lia Coelho for kindly providing of the
Cat Neurons image database. This work was financially supported by
FAPESP (processes 02/02504-01, 99/12765-2 and 96/05497-3) and to CNPQ
(process 301422/92-3).

\clearpage            
\bibliographystyle{unsrt}
\bibliography{rato}

\end{document}